\documentclass[12pt,reqno]{amsart}

\usepackage{bm}
\usepackage{amssymb}

\newtheorem{theorem}{Theorem}
\newtheorem{proposition}{Proposition}

\newtheorem{remark}{Remark}

\newcommand{\wconn}{\widetilde {\nabla}}
\newcommand{\conn}{\overline {\nabla}}

\newcommand{\wD}{\widetilde{D}}
\newcommand{\oD}{\overline{D}}

\newcommand{\s}{\mathbb{S}}

\newcommand{\T}{\mathcal{T}}

\newfont{\bb}{msbm10 at 12pt}

\def\pf{{\textit {Proof :} }}
\def\qed{\hfill{Q.E.D.}\smallskip}

\newcommand{\ls}{\setlength{\baselineskip}{12pt}
                 \setlength{\parskip}{3mm}}

\newcommand{\mysection}[1]{\section{#1}\setcounter{equation}{0}}

\newcommand{\bal}{\begin{align}}      \newcommand{\eal}{\end{align}}
\newcommand{\ba}{\begin{array}}      \newcommand{\ea}{\end{array}}
\newcommand{\bc}{\begin{center}}     \newcommand{\ec}{\end{center}}
\newcommand{\be}{\begin{enumerate}}  \newcommand{\ee}{\end{enumerate}}
\newcommand{\beq}{\begin{eqnarray}}  \newcommand{\eeq}{\end{eqnarray}}
\newcommand{\beQ}{\begin{eqnarray*}} \newcommand{\eeQ}{\end{eqnarray*}}
\newcommand{\bi}{\begin{itemize}}    \newcommand{\ei}{\end{itemize}}
\newcommand{\bt}{\begin{tabular}}    \newcommand{\et}{\end{tabular}}
\newcommand{\bdm}{\begin{displaymath}} \newcommand{\edm}{\end{displaymath}}

\begin{document}

\title{A New Quasi-local Mass and Positivity}

\address{Institute of Mathematics,
Academy of Mathematics and Systems Science, Chinese Academy of
Sciences, Beijing 100080, China}

\author{Xiao Zhang}
\thanks{Partially
supported by NSF of China(10421001), NKBRPC(2006CB805905) and the
Innovation Project of Chinese Academy of Sciences.}

\email{xzhang@amss.ac.cn}

\begin{abstract}
A new definition of quasi-local mass is proposed and its
positivity is proved under certain conditions.
\end{abstract}



\maketitle

\mysection{Introduction} \ls

Recently, Kijowski \cite{K}, Liu and Yau \cite{LY1} propose a
definition of quasi-local mass for any spacelike, topological
2-sphere with positive intrinsic (or Gauss) curvature. In
particular, Liu and Yau \cite{LY1, LY2} are able to prove its
positivity. Liu and Yau's definition arises naturally from
calculations in the recent work of Yau \cite{Y} on black holes as
a candidate for positivity. They use the original idea of Schoen
and Yau \cite{SY1, SY2} for the proof of the positive mass
conjecture to solve Jang's equation and reduce to the Riemannian
version of the positivity proved earlier by Shi and Tam \cite{ST}.

As it is well-known, there have been many attempts to define a
quasi-local mass or energy associated to a spacelike 2-surface.
Certain criterions which a definition should have were posed by
Christodoulou and Yau \cite{CY}, and Liu and Yau \cite{LY2}.
However, no known definition can satisfy all these criterions.

The definition of Kijowski, Liu and Yau belongs to the class of
definitions initiated by Brown and York \cite{BY1, BY2} for a
spacelike 2-surface which bounds a compact spacelike hypersurface
in a time orientable spacetime. Similar derivation was also
obtained by Hawking and Horowitz \cite{HH}. We refer to \cite{MST}
for a short exploiting of the idea of Brown and York. When the
spacelike 2-sphere bounds a totally geodesic spacelike
hypersurface, the definition of Kijowski, Liu and Yau reduces to
the quasi-local mass of Brown and York whose positivity was proved
in \cite{ST} for topological 2-sphere with positive Gauss
curvature and positive mean curvature.

The quasi-local energy-momentum of Brown and York has some
desirable properties. It approaches to Arnowitt-Deser-Misner (ADM)
energy-momentum in a suitable limit \cite{BY1, BLY}. However, the
decreasing monotonicity property indicates that the larger region
contains less mass (see Lemma 4.2 \cite{ST}, also \cite{LY2} for
the computation of the round sphere in time slices in the
Schwarzschild spacetime). And this breaks certain physical
intuition.

To obtain an increasing monotonicity quasi-local mass of Brown and
York's type, we propose a new definition by choosing certain spinor
norm as lapse function. The idea of using spinor norm coupled with
the mean curvature to define the quasi local mass and momentum was
told by Yau in summer 2005, which was used in \cite{WY}. If a
spacelike topological 2-sphere bounds a smooth spacelike
hypersurface (initial date set) in a spacetime whose mean curvature
does not change sign, the mean curvature of this topological sphere
satisfies apparent horizon conditions and the topological sphere can
be smoothly isometrically embedded into Euclidean 3-space whose
image has nonnegative mean curvature, then the positivity holds.
Moreover, if the initial data set has the zero second fundamental
form, the vanishing quasi-local mass implies the spacelike
topological 2-sphere bounds a domain in Euclidean 3-space. Here, we
don't assume the mean curvature of spacelike 2-sphere is strictly
positive as \cite{ST, LY1, LY2}. Thus our theorem of positivity can
apply to apparent horizons, in particular, to the round sphere in
time slices in the Schwarzschild spacetime. Indeed, we compute our
quasi-local mass in this case and find that it is an increasing
function of the radius of the round sphere.

\mysection{Preliminary} \ls

Let $(N, {\widetilde g})$ be a 4-dimensional spacetime which
satisfies the Einstein field equations
 \begin{eqnarray*}
  {\widetilde {Ric}} -\frac{\widetilde R}{2}\;{\widetilde g} = T,
 \end{eqnarray*}
where $\widetilde{Ric}$, $\widetilde R$ are Ricci curvature tensor,
scalar curvature of $\widetilde g$ respectively, and $T$ is the
energy-momentum tensor. Let $(M, g, p)$ be a smooth {\it initial
data set}, where $M$ is a spacelike (orientable) hypersurface of $N$
with the smooth induced Riemannian metric $g$ and second fundamental
form $p$. We will introduce briefly the Dirac-Witten operator along
$M$. We refer to \cite{W, PT, Z} for the basic materials.

Fix a point $p \in M$ and an orthonormal basis $\{ e _{\alpha}\}$
of $T_p N$ with $e_0$ normal and $e_i$ tangent to $M$ ($1\leq i
\leq 3$). Let $\omega ^\beta _{\;\;\alpha }$ be the connection
1-form
 \beq
\widetilde \nabla e _ \alpha = e _\beta \otimes \omega ^\beta
_{\;\;\alpha }. \label{conn1}
 \eeq
Denote
 \beQ
\omega _{\alpha \beta }= g _{\alpha \gamma }\omega ^{\gamma}
_{\;\;\beta}.
 \eeQ
Since
 \beQ 0 &=& d \big( e _\alpha ,e _\beta \big) \\
  &=& \big( \widetilde \nabla e _\alpha ,e _\beta \big)+
      \big( e _\alpha , \widetilde \nabla e _\beta \big)\\
  &=& \big(e _\gamma \otimes \omega ^\gamma _{\;\;\alpha}, e _\beta \big)
      +\big(e _\alpha, e _\delta \otimes \omega ^\delta _{\;\;\beta}
      \big),
 \eeQ
we have
 \beQ
 \omega _{\alpha \beta } = -\omega _{\beta \alpha}.
 \eeQ
Thus
 \beQ
 \omega ^0 _{\;\;i} =-\omega _{0i}=\omega _{i0}=\omega ^i _{\;\;0}.
 \eeQ
Note that for any vector $X$
 \beQ
(e _\alpha \wedge e _\beta )(X) =\tilde g (e _\alpha , X) e _\beta
-\tilde g (e _\beta , X) e _\alpha.
 \eeQ
This implies
 \beQ
(e _0 \wedge e _j) (e _0, e_1, e_2, e _3) &=& (e _0, e_1, e_2, e _3)E _{0j},\\
(e _i \wedge e _j) (e _0, e_1, e_2, e _3) &=& (e _0, e_1, e_2, e
_3)E _{ij}
 \eeQ
where $E_{0j} =(a ^0 _{\alpha \beta})$ with $a ^0 _{0j}=a ^0
_{j0}=-1$ for $j \neq 0$, $others =0$; $E_{ij} =(a _{\alpha \beta})$
with $a _{ij}=-a _{ji}=-1$ for $i < j \neq 0$, $others =0$.

Denote the $3$-vector $\Omega _0$ and $3 \times 3$ anti-symmetric
matrix $\Omega $ by
 \beQ
\Omega _0 &=& (\omega _{01}, \omega _{02},\omega _{03}),\\
\Omega   &=& (\omega _{ij}).
 \eeQ
Now (\ref{conn1}) gives that
 \beQ
\widetilde \nabla e _0 &=& -e _i \otimes \omega _{0i},\\
\widetilde \nabla e _i &=& -e _j \otimes \omega _{ij}+
                     e _0 \otimes \omega _{i0},
 \eeQ
i.e.,
 \beQ
\widetilde {\nabla}(e _0, e_1, e_2, e _3)=(e _0, e_1, e_2, e
_3)\otimes \left (
\begin{array}{cc}
         0 &  -\Omega _0  \\
         -\Omega _0 ^t  & \Omega
               \end{array}
\right).
 \eeQ
This induces a connection on the principal $SO(3,1)$ bundle with the
connection 1-form
 \beQ
\omega = -\frac{1}{2}\omega _{ij} e _i \wedge e_j +
          \omega _{0i} e _0 \wedge e _i.
 \eeQ

Denote by $\s $ the (local) spinor bundle of $N$. Since $M$ is
spin, $\s$ exists globally over $M$. This spinor bundle $\s$ is
the hypersurface spinor bundle of $M$. Let $\widetilde \nabla$ and
$\conn$ be the Levi-Civita connections of $\widetilde g$ and $g$
respectively,  the same symbols are used to denote their lifts to
the hypersurface spinor bundle. There exists a Hermitian inner
product $(\;,\;)$ on $\s$ along $M$ which is compatible with the
spin connection $\widetilde \nabla $. The Clifford multiplication
of any vector $\widetilde X$ of $N$ is symmetric with respect to
this inner product. However, this inner product is not positive
definite and there exists a positive definite Hermitian inner
product defined by $\big \langle\; ,\;\big\rangle = (e _0 \cdot\;
,\;)$ on $\s$ along $M$.

Denote $\{\sigma _a \}$ the orthonormal basis on the hypersurface
spinor bundle $\s $. Then the (local) spin connection of $N$ is
lifted from the principal $SO(3,1)$ connection as
 \beQ
\widetilde \nabla \sigma _a = -\frac{1}{4} \omega _{ij} \otimes e _i
\cdot e _j \cdot \sigma _a +\frac{1}{2} \omega _{0i}\otimes e_0
\cdot e_i \cdot \sigma _a.
 \eeQ
For spinor $\phi = u ^a \sigma _a $, vector $\widetilde X \in
\Gamma (TN)$, we have
 \beQ
\widetilde \nabla _{\widetilde X} \phi &=& du ^a (\widetilde X)
\sigma _a - \frac{u ^a }{4} \omega _{ij} (\widetilde X) e _i \cdot
e _j \cdot \sigma _a\\&& + \frac{u ^a }{2} \omega _{i0}
(\widetilde X) e _i \cdot e _0 \cdot \sigma _a .
 \eeQ
Therefore, for $X \in \Gamma (TM)$,
 \beq
\widetilde {\nabla} _X \phi=\conn _X \phi + \frac{1}{2}\omega
_{0i}(X) e_0 \cdot e _i \cdot \phi\,.
 \eeq

Define the second fundamental form of the initial data set
 \beQ
p _{ij}= \widetilde g (\widetilde \nabla _i e_0, e_j)=\omega _{j0}
(e _i).
 \eeQ
Then
 \begin{eqnarray}
\widetilde \nabla _i=\conn _i - \frac{1}{2}p _{ij} e _0\cdot e
_j\cdot. \label{two-conn}
 \end{eqnarray}
This implies that the spinor connection $\conn $ is compatible
with the positive definite inner product $\langle\;,\;\rangle$.

The Dirac-Witten operator along $M$ is defined by
 \beQ
\widetilde D =
 e _i \cdot \widetilde \nabla _i.
 \eeQ
The Dirac operator of $M$ but acting on $\s$ is defined by
 \beQ
\oD = e _i \cdot \conn _i .
 \eeQ
By (\ref{two-conn}), we obtain
 \begin{eqnarray}
\widetilde D =\oD -\frac{tr _g (p)}{2} e _0 \cdot. \label{two-D}
 \end{eqnarray}
The operator $\widetilde D$ is formally self-adjoint with respect
to $\langle \;,\; \rangle$ and its square is given by the
following Weitzenb{\"o}ck type formula
 \begin{eqnarray}
\widetilde D ^2 = \widetilde \nabla ^* \widetilde \nabla +\T
 \label{w}
 \end{eqnarray}
where
 \beQ
\T =\frac{1}{2}(T _{00} + T _{0i} e  _0 \cdot e _i\cdot),
 \eeQ
and $\wconn ^* $ is the formal adjoint of $\wconn $ with respect
to $\langle \;,\; \rangle$. If the spacetime satisfies the {\it
dominant energy condition}, then $\T$ is a nonnegative operator.

Suppose that $M$ has boundary $\Sigma $ which has finitely many
connected components $\Sigma _1, \cdots, \Sigma _l$, each of which
is a topological 2-sphere, endowed with its induced Riemannian and
spin structures. Fix a point $p \in \Sigma$ and an orthonormal
basis $\{ e _i\}$ of $T _p M$ with $e _r =e _1$ the outward normal
to $\Sigma $ and $e _a$ tangent to $\Sigma$ for $2 \leq a \leq 3$.
Let
 \beQ
h _{ab} = \langle \conn _a e _r, e _b \rangle
 \eeQ
be the component of the second fundamental form of $\Sigma $. Let
$H = tr( h )$ be its mean curvature. $\Sigma $ is a {\it
future/past apparent horizon} if
 \begin{eqnarray}
H \mp tr (p | _{\Sigma }) \geq 0 \label{horizon}
 \end{eqnarray}
holds on $\Sigma $. When $\Sigma $ has multi-components, we require
that (\ref{horizon}) holds (with the same sign) on each $\Sigma _i$.

Denote by $\nabla $ the lift of the Levi-Civita connection of
$\Sigma $ to the spinor bundle $\s | _{\Sigma}$. Then
 \begin{eqnarray}
\conn _a = \nabla _a+\frac{1}{2} h _{ab} e _r\cdot e _ b \cdot\;.
\label{two-conn2}
 \end{eqnarray}
The connection $\nabla $ is also compatible with the metric
$\langle \;,\; \rangle$ on the boundary $\Sigma$. Let
 \beQ
D=e _a \cdot \nabla _a
 \eeQ
be the Dirac operator of $\Sigma$ but acting on $\s | _{\Sigma}$.
From (\ref{two-conn}), (\ref{two-conn2}), we obtain
 \begin{eqnarray}
\widetilde \nabla _a =\nabla _a +\frac{1}{2}h _{ab} e _r \cdot e
_b \cdot -\frac{1}{2}p _{aj} e _0 \cdot e _j \cdot.
\label{two-conn3}
 \end{eqnarray}
Therefore (\ref{w}) gives rise to
 \begin{eqnarray}
\int _M  |\widetilde \nabla \phi |^2 + \langle \phi, \T \phi
\rangle - |\widetilde D \phi |^2
 &=& \int _{\Sigma} \langle \phi, (e _r\cdot D -\frac{H}{2})\phi \rangle
 \nonumber\\
& &+\frac{1}{2}\int _{\Sigma }\langle \phi, tr (p | _{\Sigma }) e
_0 \cdot e _ r\cdot \phi \rangle \nonumber\\
 & &-\frac{1}{2}\int _{\Sigma }\langle \phi, p _{ar} e _0 \cdot e _a
\cdot \phi\rangle . \label{w2}
 \end{eqnarray}

We will prove the following existences. Let
 \beQ
P _{\pm}=\frac{1}{2}(Id \pm e _0 \cdot e _r \cdot )
 \eeQ
be the projective operators on $\s | _{\Sigma}$. The local
boundary conditions
 \beQ
P _{\pm} \phi  | _{\Sigma} = 0
 \eeQ
ensure that $\widetilde D$ is a formally self-adjoint elliptic
operator \cite{GHHP}.

\begin{proposition}
Let $(N, {\widetilde g})$ be a spacetime which satisfies the
dominant energy condition. Let $(M, g, p)$ be a smooth initial
data with the boundary $\Sigma$ which has finitely many
multi-components $\Sigma _i $, each of which is topological
2-sphere.

(i) If $tr _g (p) \geq 0$ and $\Sigma$ is a past apparent horizon,
then the following Dirac-Witten equation has a unique smooth
solution $\phi \in \Gamma (\s)$
 \begin{eqnarray}
 \left\{ \begin{array}{ccccc}
       \wD \phi &=&  0 &in & M\\
  P _{+} \phi &=& P _{+}  \phi _0 & on &\Sigma _{i _0} \\
  P _{+} \phi &=& 0 & on &\Sigma _{i}\;(i \neq i _0)\\
    \end{array} \right . \label{existence1}
 \end{eqnarray}
for any given $\phi _0 \in \Gamma (\s \big| _{\Sigma})$ and for
fixed $i _0 $;

(ii) If $tr _g (p) \leq 0$ and $\Sigma$ is a future apparent
horizon, then the following Dirac-Witten equation has a unique
smooth solution $\phi \in \Gamma (\s)$
 \begin{eqnarray}
 \left\{ \begin{array}{ccccc}
       \wD \phi &=&  0 &in & M\\
  P _{-} \phi &=& P _{-}  \phi _0 & on &\Sigma _{i _0} \\
  P _{-} \phi &=& 0 & on &\Sigma _{i}\;(i \neq i _0)\\
    \end{array} \right . \label{existence2}
 \end{eqnarray}
for any given $\phi _0 \in \Gamma (\s \big| _{\Sigma})$ and for
fixed $i _0 $.
\end{proposition}
\pf To establish the existences of (\ref{existence1}) and
(\ref{existence2}), we only need to show that any solution $\wD
\phi =0$ is trivial when $\phi _0 =0$. By (\ref{two-D}), we obtain
 \beQ
\wD (e _0 \cdot \phi ) = -tr _g (p) \phi.
 \eeQ
Therefore,
 \begin{eqnarray*}
\int _{\Sigma } \langle e _r \cdot \phi , e _0 \cdot \phi \rangle
&=& \int _M \langle \wD \phi, e _0 \cdot \phi \rangle - \langle
\phi, \wD (e _0 \cdot \phi)\rangle\\
&=&\int _M tr _g (p) | \phi | ^2.
 \end{eqnarray*}
If $\Sigma $ is a past apparent horizon and $tr _g (p) \geq 0$, or
$\Sigma $ is a future apparent horizon and $tr _g (p) \leq 0$,
then $\phi | _{\Sigma } =0$. Note that, under the local boundary
conditions, (\ref{w2}) becomes
 \beq
\int _M |\widetilde \nabla \phi |^2 + \langle \phi, \T \cdot \phi
\rangle -|\widetilde D \phi |^2=-\frac{1}{2}\int _{\Sigma}\big[ H
\pm tr (p| _{\Sigma}) \big]|\phi | ^2. \label{w3}
 \eeq
This implies $\wconn _i \phi =0$ on $M$. Thus
 \beQ
|d | \phi | ^2| \leq 2 | \conn \phi | | \phi |\leq |p| | \phi | ^2
\leq C_p | \phi | ^2,
 \eeQ
where $C_p =max |p|$. If there exists a point $x_0 \in M$ such that
$\phi (x _0) \neq 0$, then
 \beQ
| \phi | (x) \geq | \phi | (x _0) e ^{-C _p \rho _{x _0} (x)}.
 \eeQ
Taking $x$ to the boundary will gives rise to a contradiction.
Therefore $\phi =0$ on $M$ and the existences of
(\ref{existence1}) and (\ref{existence2}) follow. \qed

\mysection{Definition and Positivity}
\ls

Let $(N, {\widetilde g})$ be a spacetime which satisfies the
dominant energy condition. Let $(M, g, p)$ be a smooth initial
data set with the boundary $\Sigma$ which has finitely many
multi-components $\Sigma _i $, each of which is topological
2-sphere. Let $e _r$ be the unit vector outward normal to $\Sigma
$, $h _{ij}$ and $H$ be the second fundamental form and the mean
curvature of $\Sigma$ respectively.

Suppose that some $\Sigma _{i_0} $ can be smoothly isometrically
embedded into $\mathbb{R} ^3 $ in the Minkowski spacetime
$\mathbb{R} ^{3,1}$ and denote by $\aleph$ the isometric embedding.
(It exists if $\Sigma _{i_0} $ has positive Gauss curvature.) Let
$\breve{\Sigma}$ be the image of $\Sigma $ under the map $\aleph$.
Let $\breve{e}_r$ the unit vector outward normal to $\breve{\Sigma}
$ and $\breve{h}_{ij}$, $\breve{H}$ are the second fundamental
forms, the mean curvature of $\breve{\Sigma}$ respectively. Denote
by
 \beQ
H _0 = \breve{H} \circ \aleph
 \eeQ
the pullback to $\Sigma$.

The isometric embedding $\aleph$ also induces an isometry between
the (intrinsic) spinor bundles of $\Sigma _{i _0}$ and
$\breve{\Sigma} _{i _0}$ together with their Dirac operators which
are isomorphic to $e _r \cdot D$ and $\breve{e} _r \cdot \breve{D}$
respectively. This isometry can be extended to an isometry over the
complex 2-dimensional sub-bundles of their hypersurface spinor
bundles. Denote by ${\breve{\s} ^{\breve{\Sigma} _{i _0}}}$ this
sub-bundle of $\breve{\s} | _{\breve{\Sigma} _{i _0}}$. Let
$\breve{\phi}$ be a constant section of ${\breve{\s}
^{\breve{\Sigma} _{i _0}}}$ and denote
 \beQ
\phi _0 = \breve{\phi} \circ \aleph.
 \eeQ
Denote by $\breve{\Xi}$ the set of all these constant spinors
$\breve{\phi}$ with the unit norm. This set is isometric to $S ^3$.

Suppose that one of the following conditions holds on $M$

$(i)$ $tr _g(p) \geq 0$, $H | _{\Sigma _i} + tr (p | _{\Sigma _i})
\geq 0$ for all $i$;

$(ii)$ $tr _g(p) \leq 0$, $H | _{\Sigma _i} - tr (p | _{\Sigma _i})
\geq 0$ for all $i$.

Let $\phi $ be the unique solution of (\ref{existence1}) or
(\ref{existence2}) for some $\breve{\phi} \in \breve{\Xi}$. Denote
 \beq
m(\Sigma _{i _0}, \breve{\phi})&=&\frac{1}{8\pi}\Re \int _{\Sigma
_{i_0}}
\Big[(H _0 -H ) | \phi | ^2 \nonumber\\
& & + tr(p |_{\Sigma _{i_0}}) \langle \phi, e_0 \cdot e _r \cdot
\phi \rangle  \nonumber\\
& & - p _{ar} \langle \phi, e_0 \cdot e _a \cdot \phi \rangle
\Big]. \label{pre-local-mass}
 \eeq
The {\it quasi local mass of $\Sigma _{i _0}$} is defined as
 \beq
m(\Sigma _{i _0})=\min _{\breve{\Xi}}m(\Sigma _{i _0},
\breve{\phi}). \label{local-mass}
 \eeq
If all $\Sigma _i $ can be smoothly isometrically embedded into
$\mathbb{R} ^3 $ in the Minkowski spacetime $\mathbb{R} ^{3,1}$, we
define the {\it quasi local mass of $\Sigma $} as
 \beq
m(\Sigma )=\sum _i m(\Sigma _{i}). \label{total-local-mass}
 \eeq

It is obvious that $m(\Sigma _{i _0})$ vanishes if $\Sigma _{i _0}$
is an embedding 2-sphere in $\mathbb{R} ^3 $. Let $E, P_1, P_2, P_3$
be the ADM energy-momentum of an asymptotically flat spacetime
\cite{SY2, W, PT, Z}. Take $\Sigma _{i_0}$ to be a sphere in an end.
Then
 \beQ
\lim _{r \rightarrow \infty} m(\Sigma _{i _0})&=&\min
_{\breve{\Xi}}\langle \breve{\phi}, E \breve{\phi} + P _k {dx} _0
\cdot {dx} _k \cdot \breve{\phi} \rangle \\
&=& E -|P|.
 \eeQ

 \begin{theorem}\label{thm1}
Let $(N, {\widetilde g})$ be a spacetime which satisfies the
dominant energy condition. Let $(M, g, p)$ be a smooth initial
data set with the boundary $\Sigma$ which has finitely many
multi-components $\Sigma _i $, each of which is topological
2-sphere. Suppose that some $\Sigma _{i_0} $ can be smoothly
isometrically embedded into $\mathbb{R} ^3 $ whose image in
$\mathbb{R} ^3 $ has nonnegative mean curvature. If either
condition $(i)$ or condition $(ii)$ holds, then
 \begin{enumerate}
 \item $m(\Sigma _{i _0}) \geq 0$;
 \item that $m(\Sigma _{i _0})=0$ implies the energy-momentum
of spacetime satisfies
 \beQ
T_{00} = |f| |\phi | ^2, \;\;\;\;T _{0i} = f \langle \phi, e_0 \cdot
e _i \cdot \phi \rangle
 \eeQ
along $M$, where $f$ is a real function, $\phi $ is the unique
solution of (\ref{existence1}) or (\ref{existence2}) for some
$\breve{\phi} \in \breve{\Xi}$.
 \item Furthermore, if $p _{ij}=0$, then
$m(\Sigma _{i _0})=0$ implies that $M$ is flat with connected
boundary.
 \end{enumerate}
 \end{theorem}
 \pf Note that (\ref{two-conn3}) implies
 \beQ
 \breve{\nabla} _a
\breve{\phi} +\frac{1}{2} \breve{h}
 _{ab} \breve{e} _r \cdot \breve{e} _b \cdot
\breve{\phi} =0
 \eeQ
on $\breve{\Sigma}$. Then
 \beQ
e _r \cdot D \phi _0=\frac{H_0}{2} \phi _0
 \eeQ
on $\Sigma$. Denote $\phi _0 ^\pm = P _{\pm} \phi _0$. Since $e _r
\cdot D \circ P _{\pm} =P _{\mp} \circ e _r \cdot D$, we have
 \beQ
e _r \cdot D \phi _0 ^{\pm} =\frac{H_0}{2}\phi _0 ^{\mp}.
 \eeQ
Therefore, using the self-adjointness of $e _r \cdot D$, we obtain
  \beQ
\int _{\Sigma} H _0 |\phi _0 ^+ | ^2 =\int _{\Sigma} H _0 |\phi _0
^- | ^2.
 \eeQ
Denote $\phi ^{\pm} =P _{\pm} \phi$. Then $\int _{\Sigma} \langle
\phi ^+, \phi ^- \rangle =0$. In case $(i)$, we have $\phi ^+
=\phi _0 ^+$. Thus
 \beQ
\int _{\Sigma} \langle \phi,e _r \cdot D \phi\rangle &=& 2 \Re
\int _{\Sigma} \langle \phi ^-,e _r \cdot D \phi ^+ \rangle \\
&=& 2 \Re
\int _{\Sigma} \langle \phi ^-,e _r \cdot D \phi ^+ _0 \rangle \\
&=& \Re \int _{\Sigma} \langle \phi ^-, H _0 \phi _0 ^- \rangle \\
&\leq &\frac{1}{2} \int _{\Sigma}  H _0  \big(|\phi ^- | ^2+|\phi _0 ^- | ^2 \big)\\
&= &\frac{1}{2} \int _{\Sigma}  H _0  \big(|\phi ^- | ^2+|\phi _0 ^+ | ^2 \big)\\
&=&\frac{1}{2} \int _{\Sigma}  H _0  |\phi | ^2.
 \eeQ
Therefore $m(\Sigma _{i _0}, \breve{\phi}) \geq 0$ which implies
the positivity. Same argument applies to the case $(ii)$. Thus the
proof of the first part is complete.

For the second part, if $m(\Sigma _{i _0})=0$, then there exists
$\breve{\phi}$ such that $m(\Sigma _{i _0}, \breve{\phi})=0$. Let
$\phi$ be the unique solution of (\ref{existence1}) or
(\ref{existence2}) corresponding to the boundary value
$\breve{\phi}$. Then (\ref{w2}) implies $\wconn _i \phi =0$ on $M$.
This gives that $|\phi |$ is nonzero and
 \beQ
\langle \phi, (T _{00} +T _{0i} e _0 \cdot e _i)\phi \rangle =0
 \eeQ
along $M$. Therefore, the dominant energy condition and Lemma 1
\cite{HZ2} solve $T_{00}$ and $T _{0i}$. Furthermore, if $p_{ij}=0$
in this case, then there exists a parallel spinor on $M$. Thus $M$
is Ricci flat, hence flat since it is 3-dimensional. As each
component of its boundary has nonnegative mean curvature, the
boundary must be connected. \qed

 \begin{remark}
The vanishing quasi local mass in Theorem 1 gives rise to one spinor
which is parallel along $M$ and it does not imply the Ricci flatness
of the spacetime along $M$. This is similar to Dougan-Mason's quasi
local mass \cite{DM}. There is a standard model for 4-dimensional
spacetimes, i.e., pp-manifolds, admitting one parallel spinor,
 \beQ
\tilde g_{f} =-2 dx_1 dx_2 +f(x_2, x_3, x_4) dx _2 ^2 +dx _3 ^2 +dx
_4 ^2,
 \eeQ
where $f$ is a smooth function (e.g.,\cite{B}). $\tilde g_{f}$ is
Ricci flat if and only if $f$ satisfies
 \beQ
\frac{\partial ^2 f}{\partial x _3 ^2} + \frac{\partial ^2
f}{\partial x _4 ^2}=0.
 \eeQ
 \end{remark}
 \begin{remark}
The second fundamental form of the time slice in the Kerr spacetime
is nontrivial and falls off as $\frac{1}{r^3}$ (e.g. \cite{Z1}).
Thus, for the regular initial data set which is Kerr at infinity
constructed in \cite{CS}, the ``quasi-local momentum" of a bounded
domain with sufficiently large radius may be nontrivial although the
total ADM linear momentum is zero.
 \end{remark}

We shall use null normals to express $m(\Sigma _{i _0},
\breve{\phi})$. At each point of $\Sigma _{i_0}$, we choose two
null normal vectors $n _{+}$ and $n _{-}$ such that $\widetilde g
(n _{+}, n _{-})=-1$. We also choose the local frame $e _2$, $e
_3$ of $\Sigma $. Denote by ${\bf H} \equiv -\wconn _{e _2} e _2
-\wconn _{e _3} e _3$ the mean curvature vector of $\Sigma $. Let
 \begin{eqnarray*}
2 \rho = \widetilde g ({\bf H}, n _{+}), \;2 \mu = \widetilde g
({\bf H}, n _{-}), \;\varpi _a =-\widetilde g (\wconn _a n _{+}, n
_{-})
 \end{eqnarray*}
where $a=2,3$. A straightforward computation yields
 \begin{eqnarray*}
tr(p |_{\Sigma _{i _0}}) =\sqrt{2}(\rho -\mu),\;p _{ar} =\varpi
_a.
 \end{eqnarray*}
Thus
 \begin{eqnarray*}
m(\Sigma _{i _0}, \breve{\phi})&=&\frac{1}{8\pi}\Re \int _{\Sigma
_{i_0}} \Big[(H _0 -H ) | \phi | ^2 \nonumber\\
& & -\frac{\rho -\mu}{\sqrt{2}} \langle \phi, (n _{+} \cdot n
_{-}  \cdot -  n _{-}  \cdot n _{+}  \cdot )\phi \rangle\\
& & -\frac{\varpi _{a}}{\sqrt{2}} \langle \phi, (n _{+} + n _{-})
\cdot e _a \cdot \phi \rangle \Big].
 \end{eqnarray*}

This formula suggests a way to define the quasi-local mass for
spacelike 2-spheres $\Sigma $ using geometric data of their
embedding into a 4-dimensional spacetime. However, it is still
unclear what the correct choice of spinor $\phi$ is. In the cases
that $\Sigma $ is a past apparent horizon and encloses a spacelike
hypersurface $M ^+$ with nonnegative mean curvature, or $\Sigma $
is a future apparent horizon and encloses a spacelike hypersurface
$M ^-$ with nonpositive mean curvature, then we can choose spinor
$\phi $ as the solution (\ref{existence1}) or (\ref{existence2}).
Denote $m(\Sigma _{i _0}, \breve{\phi})$ as $m(\Sigma,
\breve{\phi}, M ^\pm)$ for a chosen spacelike hypersurface $M
^\pm$. Then we can define the quasi-local mass of $\Sigma $ as
\begin{eqnarray*}
m(\Sigma )=\inf _{\{M ^\pm \}} \min _{\breve{\Xi}}m(\Sigma ,
\breve{\phi}, M ^\pm)
 \end{eqnarray*}
where $\{M ^\pm \}$ is the set of all required spacelike
hypersurfaces enclosed by $\Sigma$. If $\Sigma $ satisfies all
conditions mentioned above, we can still prove $m(\Sigma )$ is
nonnegative.

\mysection{Schwarzschild spacetime} \ls

We conjecture that $m(\Sigma )$ has increasing monotonicity. We
verify that it is indeed the case for the round spheres in
time-slices $M$ of the exterior Schwarzschild spacetime. For this
$M$, the metric is
 \beQ
g_{Sch} =\big(1-\frac{2m}{r} \big)^{-1} dr ^2 +r ^2 \big( d \theta
^2 +\sin ^2 \theta d \psi ^2 \big)
 \eeQ
$(r \geq 2m)$ and the second fundamental form vanish, i.e., $p
_{ij} =0$. Using the isotropic coordinate $r =\rho
\big(1+\frac{m}{2\rho}\big) ^2$ $(\rho \geq \frac{m}{2})$, $g
_{Sch}$ can be written as
 \beQ
g_{Sch} =\big(1+\frac{m}{2 \rho} \big)^{4} \big[d\rho ^2 +\rho ^2
\big( d \theta ^2 +\sin ^2 \theta d \psi ^2 \big)\big]\equiv u ^4
g _{flat}
 \eeQ
where $u=1+\frac{m}{2 \rho}$. $g _{Sch}$ has an apparent horizon
$\Sigma =\{\rho =\frac{m}{2}\}$ which is a minimal surface.

We briefly discuss the Dirac operator for conformal metrics. We
refer to \cite{HZ} and references therein for detail. Denote by
$G_u$ the isometry between ${\rm SO}_{g_{flat}} $ and ${\rm SO}
_{g_{Sch}}$ given by the above conformal change of metric. The
isometry $G _u$ induces an isometry between the ${\rm Spin}_3$
principal bundles as well as an isometry between their spinor
bundles $S$ and $\widetilde S $ ($\equiv G_u S$). For any sections
$\phi, \psi$ of the spinor bundle $S$, we denote $ \tilde \phi = G
_u \phi$, and $\tilde \psi = G _u \psi$ the corresponding sections
of $\widetilde  S$. If $(\; , \;)_{g _{flat}}$ and $(\; , \;)_{g
_{Sch}}$ denote respectively the natural Hermitian metrics on $S$
and $\widetilde S$, then
 \beQ
(\phi,\psi)_{g _{flat}}=(\tilde{\phi},\tilde{\psi})_{g _{Sch}}.
 \eeQ

The Clifford multiplication on $\widetilde S$ is given by
 \beQ
\tilde{e _i} \;\tilde{\cdot} \; \tilde{\phi} = {\widetilde {e _i
\cdot\phi}}.
 \eeQ
Dirac operators with respect to the two metric have the relation
 \beQ
{\overline D} _{Sch} \big( u ^{-1} \tilde{\phi}\big) = u
^{-3}{\widetilde {{\overline D}_{flat} \phi}}.
 \eeQ
For any covariant constant spinor $\breve{\phi}$ with unit norm,
$u ^{-1} \widetilde{\breve{\phi}} $ is a solution of $\overline D
_{Sch}$. Moreover, $| \widetilde {\breve{\phi}} | ^2 _{Sch} = |
\breve{ \phi} | ^2 _{flat} =1$.

For any round sphere $B _{\rho}$ in $M$ with radius $\rho$, the
Brown-York quasi-local mass is (e.g. \cite{LY2})
 \beQ
m _{BY} (B _\rho) &=& \frac{1}{8\pi} \int _{S _\rho} (H _0 -H )\\
&=&r \big(1- \sqrt{1-\frac{2m}{r}} \big)\\
&=&m \big(1+\frac{m}{2\rho}\big).
 \eeQ
Since $H _0 >0$, the new quasi-local mass is
 \beQ
m (B _\rho ) &=& \min _{\breve{\Xi}} \frac{1}{8 \pi}\int _{S _\rho}
(H _0 -H )
| u ^{-1} \widetilde{\breve{\phi}}  | ^2 _{Sch}\\
&=&r \Big(1-\sqrt{1-\frac{2m}{r}}\Big) u ^{-2} \\
&=&\frac{m}{1 +\frac{m}{2 \rho}}.
 \eeQ
Therefore $m (B _\rho )$ is increasing and
 \beQ
\lim _{\rho \rightarrow \infty} m \big(B _\rho \big) =m =E _{ADM}.
 \eeQ

{\footnotesize {\it Acknowledgements.} The author is indebted to
M.T.Wang and S.T.Yau for sharing their idea and some valuable
conversations.}

\end{document}